\documentclass[epj-spec]{svjour}
\usepackage{graphics}
\begin{document}
\title{Stochastic resonance in finite arrays of bistable elements with local coupling }
\author{Manuel Morillo \inst{1}, Jos\'e G\'omez Ord\'o\~nez \inst{1}, Jos\'e M. Casado \inst{1}, Jes\'us
Casado-Pascual \inst{1} and David Cubero\inst{2}
\thanks{We acknowledge the support of the Ministerio de Ciencia e Innovaci\'on of Spain (FIS2008-4120(MM,JMC,JGO) and FIS2008-02873(JC-P,DC))
 and the Junta de Andaluc\'{\i}a.}
}                     
%
%
\institute{F\'{\i}sica Te\'orica. Facultad de F\'{\i}sica.
Universidad de Sevilla. Apartado de Correos 1065. Sevilla 41080.
Spain \and Departamento de F\'{\i}sica Aplicada I, EUP, Universidad de Sevilla, Calle Virgen de \'Africa 7, 41011 Sevilla, Spain}
%
\date{Received: date / Revised version: date}
%
\abstract{ In this article, we investigate the stochastic resonance
(SR) effect in a finite array of noisy bistable systems with
nearest-neighbor coupling driven by a weak time-periodic driving
force. The array is characterized by a collective variable. By means
of numerical simulations, the signal-to-noise ratio (SNR) and the
gain are estimated as functions of the noise and the interaction
coupling strength. A strong enhancement of the SR phenomenon for
this collective variable in comparison with SR in single unit
bistable systems is observed. Gains larger than unity are obtained
for some parameter values and multi- frequency driving forces,
indicating that the system is operating in a non-linear regime
albeit the smallness of the driving amplitude. The large SNR values
observed are basically due to the fact that the output fluctuations
are small and short lived, in comparison with their typical values
in a linear regime. A non-monotonic behavior of the  SNR with the
coupling strength is also obtained. }



%
\authorrunning{Manuel Morillo et al.}
\titlerunning{Stochastic resonance in finite arrays}

\maketitle
\section{Introduction}
\label{intro} The study of the response of complex systems formed by
\emph{coupled } nonlinear noisy elements subject to the action of a
time-periodic force brings into the picture a new ingredient with
respect to the response of a single unit: the effect of the
interactions. The interplay of noise, intrinsic nonlinear dynamics,
driving forces and interactions render the response a very rich
stochastic process. Certain aspects of the response are associated
with what is called  stochastic resonance (SR) \cite{RMP,chaos}, a
phenomenon of interest to scientific fields other than Physics (as
discussed, for instance, in \cite{cpc}). SR is characterized by the
non-monotonic behavior of some quantifiers with the noise strength.
An enhancement of SR effects with respect to those observed in
single units have been reported in several papers
\cite{jung,schm,our,neiman,lind95}. More recently
\cite{us06,allus07,david}, we have analyzed the \emph{collective}
response of a finite set of globally coupled bistable systems. We
demonstrated that the response of the collective variable shows SR
effects which are indeed much enhanced with respect to those in
single bistable units. Actually, in those arrays,  SR gains larger
than unity were observed, we believe for the first time, for
\emph{subthreshold sinusoidal} driving forces. Those findings
indicated that the arrays were indeed operating in nonlinear
regimes. Similar results have been reported for single bistable
units driven by a suprathreshold sinusoidal force
\cite{Hanggi-Inchiosa}, by subthreshold multi-frequency forces
\cite{Loerincz,Gingl,us03,C-P1,C-P2,C-P3}, or by a subthreshold
sinusoidal force in the presence of a strong, high-frequency
monochromatic signal \cite{C-P4}. Nevertheless, to the best of our
knowledge, gains larger than unity have never been
 observed when the single bistable unit is driven
by {\em just} a subthreshold sinusoidal force.

In a later work \cite{us08}, the response of the collective variable
of a finite array of globally coupled units to a rather \emph{weak}
driving force was analyzed. Two aspects of the response were
studied: SR and noise induced phase frequency synchronization. The
fact that SR effects were greatly amplified and that very good phase
frequency synchronization was found in the response to a weak driver
led us to indicate that the arrays were indeed operating in
nonlinear regimes.

In this work, we extend the results of the previous work to other
topologies of the array. In \cite{lind95}, arrays of nonlinear
bistable units with nearest neighbor coupling are considered. In
that work the response of a single oscillator is found to be
enhanced due to the coupling to the other units in the chain. In
\cite{schi}, the authors consider an array of driven spins with
Glauber dynamics and nearest neighbor interaction. They study not
only the response of a single spin, but also the behavior of the
global magnetization of the array. Their analysis of the system
response is based on the linear response approximation. In both
papers \cite{lind95,schi}, the coupling strength among the elements
of the chain is suggested as a suitable parameter besides the noise
strength to efficiently perform desired operations. In this work,
rather than dealing with a globally coupled network, we will
consider as in \cite{lind95} that the bistable units have nearest
neighbor interactions. We will still be interested in \emph{finite}
arrays. While in \cite{lind95} the authors focus on the response of
a single oscillator, we will concentrate here as in \cite{us06,schi}
on the behavior of a collective or global variable characterizing
the entire array rather than on a single individual. By contrast
with the range of parameters analyzed in \cite{schi}, we will be
dealing here with situations where a linear response approximation
is not valid.

The rest of the paper is as follows. In Section \ref{sec:1}, we
introduce the model system, fix notation and indicate the relevant
variables that we will use to quantify the SR effect. In Section
\ref{sec:2}, we present the main results of our numerical
simulations. In Section \ref{sec:3} we comment on the main
conclusions of our paper.

\section{The model system}
\label{sec:1}
We consider a set of $N$ identical bistable elements characterized
by the variables $x_i(t)\, (i=1,\ldots,N)$ with nearest neighbor
interactions. The dynamics is given by stochastic evolution
equations (in dimensionless form) of the type
\begin{equation}
\label{lang}
\dot{x}_i(t)=x_i(t)-x_i^3(t)+\frac{\theta}{2}[x_{i-1}(t)+x_{i+1}(t)-2x_i(t)]+\sqrt{2D}\xi_i(t)+F(t),
\label{eq:lang}
\end{equation}
subject to the conditions $x_{N+1}(t)=x_1(t),\;x_0(t)=x_N(t)$. The
external driving force is periodic in time with period $T$, i. e.,
$F(t)=F(t+T)$. The term $\xi_i(t)$ represents a white noise with
zero average and $\left\langle \xi_i(t) \xi_j(s) \right\rangle =
\delta_{ij}\delta (t-s)$.  In the $N \rightarrow \infty $ limit,
this model becomes the classical $\phi^4$ model analyzed in
\cite{marche}. In the absence of driving, it has also been used to
model the dynamics of voltage pulses along myelinated nerves
\cite{scott}. In the context of the ratchet effect, a similar model
has been studied in \cite{denisov}.

We define a collective variable $S(t)$ as
\begin{equation}
S(t)=\frac 1N \sum_{j=1}^{N} x_j(t).
\end{equation}
We will concentrate on the SR effects associated with the collective
variable when the system size, $N$, is kept finite and the amplitude
of the driving term is small. By small we mean that when the driving
force acts on a single isolated unit, SR is well described by linear
response theory.

We will use the  signal-to-noise ratio (SNR) of the collective
behavior as the quantifier of the SR effects. Its definition
requires the evaluation of  the one-time correlation function
defined as
\begin{equation}
L(\tau)=\frac 1T \int_0^T \,dt\; \langle
S(t)S(t+\tau)\rangle_{\infty}.
\end{equation}
The notation $\langle \ldots \rangle$ indicates an average over the
noise realizations and the subindex $\infty$ indicates the long time
limit of the noise average, i. e., its value after waiting for $t$
long enough for the transients to die out. As indicated in our
previous work \cite{us06}, we can write
\begin{equation}
L(\tau)=L_\mathrm{coh}(\tau)+L_\mathrm{incoh}(\tau),
\end{equation}
where the coherent part, $L_\mathrm{coh}(\tau)$,
\begin{equation}\label{}
L_\mathrm{coh}(\tau)=\frac 1T \int_0^T \,dt\; \langle
S(t)\rangle_{\infty} \langle S(t+\tau)\rangle_{\infty} ,
\end{equation}
is periodic in $\tau$ with the period of the driving force, while
the incoherent part, $L_\mathrm{incoh}(\tau)$ arising from the
fluctuations of the output $S(t)$ around its average value, decays
to zero as $\tau$ increases. The output SNR, $R_\mathrm{out}$, is
\begin{equation}
\label{snr} R_\mathrm{out} =\lim_{\epsilon \rightarrow 0^+}\frac {
\int_{\Omega-\epsilon}^{\Omega+\epsilon} d\omega\;
\tilde{L}(\omega)}{\tilde{L}_\mathrm{incoh}(\Omega)}=\frac {
\tilde{L}_\mathrm{coh}(\Omega)}{\tilde{L}_\mathrm{incoh}(\Omega)} ,
\end{equation}
where $\Omega =2\pi/T$ is the fundamental frequency of the
driving force $F(t)$, $\tilde{L}_\mathrm{coh}(\Omega)$ is the
corresponding Fourier coefficient in the Fourier series expansion of
$L_\mathrm{coh}(\tau)$, and $\tilde{L}_\mathrm{incoh}(\Omega)$ is
the Fourier transform at frequency $\Omega$ of
$L_\mathrm{incoh}(\tau)$.

 We will also discuss the SR gain, $G$, defined as \cite{us06}
\begin{equation}
\label{gain} G=\frac{R_\mathrm{out}}{R_\mathrm{inp}},
\end{equation}
where $R_\mathrm{inp} $ is the SNR of the random input
process formed by the arithmetic mean of the individual noise terms
$\xi_i(t)$ plus the deterministic driving force $F(t)$, namely,
$F(t)+\xi(t)$ with $\xi(t)=N^{-1}\sum_{i=1}^N\xi_i(t)$.
The gain can be seen as a dimensionless parameter that compares the
output SNR to that of the input and, in this sense, it measures the
quality of the output relative to the input.

\section{Results} \label{sec:2} Following the numerical procedure
detailed in our previous works \cite{us03,us08}, we have estimated
the coherent and incoherent part of the collective correlation
function $L(\tau)$, by integrating the Langevin equations, Eq.\
(\ref{lang}) and averaging over several thousand noise realizations.
With this information, we evaluate numerically the integrals
defining the Fourier coefficients at the driving frequency and,
using Eqs.\ (\ref{snr}) and (\ref{gain}), the collective SNR and
gain for a wide range of parameter values.  We have used a weak
driving rectangular force given by
\begin{equation}
\label{force} F(t)=(-1)^{n(t)} A, \label{Eq009}
\end{equation}
where $n(t)=\lfloor 2\, t/T \rfloor$, $\lfloor z \rfloor$ is the
floor function of $z$, i.e., the greatest integer less than or equal
to $z$. We have also considered the case of a sinusoidal driving
term $ A\sin(\Omega t)$. In both cases, the force amplitude will be
taken to be $A=0.1$, much smaller than the barrier height of an
isolated bistable unit while the fundamental frequency will be
$\Omega=0.01$.

\begin{figure}
\resizebox{0.4\textwidth}{!}
  {\includegraphics{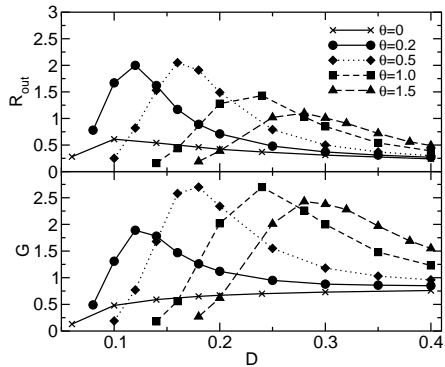}}
\caption{The collective SNR, $R_\mathrm {out}$, (upper panel) and
the collective gain, $G$, (lower panel) vs. the noise strength $D$
for an array of $N=10$ bistable units driven by a rectangular
driving force with amplitude $A=0.1$, fundamental frequency
$\Omega=0.01$ and several values of the coupling parameter:
$\theta=0$ (crosses), 0.2 (circles), 0.5 (diamonds), 1.0 (squares),
and 1.5 (triangles). Lines are a guide to the eye.}
\label{fig:1-2}       
\end{figure}

As noticed in \cite{us06}, in the case of noninteracting units
($\theta=0$), the SNR of the collective output is $N$ times larger
than that of a isolated unit driven by the same force. Nonetheless,
as discussed in \cite{us06}, the gain associated with the collective
output is just the same as the one of a single, isolated, unit.
Thus, for the weak forces that we are considering here, the
collective gain does not exceed unity, in agreement with the
predictions of the linear response theory \cite{us03,Dyk,dewbia95}.
The introduction of interactions between the bistable units
drastically changes this picture and the enhancement of the SNR
leads to the possibility of observing gains larger than unity for
the weak driving forces considered. These facts can be observed in
Fig. \ref{fig:1-2}.

The non-monotonic behavior of the SNR for the collective variable
with the noise strength observed in  the upper panel of Fig.
\ref{fig:1-2} is indicative of the SR phenomenon. In  the upper
panel of Fig. \ref{fig:1-2}, we depict the behavior of the global
$R_\mathrm {out}$ for an array of $N=10$ identical particles with
nearest neighbor coupling. The values of the interaction parameter
range from small values ($\theta=0.2$) to rather large ones
($\theta=1.5$). The $R_\mathrm {out}$ peak value depends on the
coupling strength in such a way that as $\theta$ is increased, the
noise value at which $R_\mathrm{out}$ reaches its maximum is shifted
slightly to higher values.

\begin{figure}
\resizebox{0.4\textwidth}{!}
  {\includegraphics{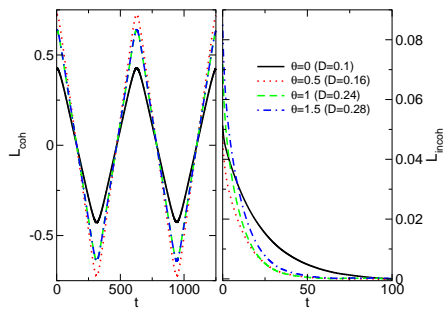}}
\caption{The coherent part, $L_\mathrm{coh}(t)$, (left panel) and
the incoherent part, $L_\mathrm{incoh}(t)$, (right panel)  of the
correlation function of the collective variable for several values
of the coupling parameter: $\theta=0$ (solid line), 0.5 (dotted),
1.0 (dashed), and 1.5 (dot-dashed), corresponding to the noise
strength values $D=0.1$, 0.16, 0.24, and 0.28, respectively. These
noise values correspond to the peaks observed in $R_\mathrm{out}$ in
 the upper panel of Fig.~\ref{fig:1-2}. Other parameter values:
$N=10$, amplitude $A=0.1$ and fundamental frequency $\Omega=0.01$. }
\label{fig:3-4}       
\end{figure}

It is interesting to compare the time behavior of the coherent,
$L_\mathrm{coh}(t)$, and incoherent, $L_\mathrm{incoh}(t)$, parts of
the collective correlation function. In the left panel of Fig.
\ref{fig:3-4}, we depict the time behavior of the coherent part for
several values of the interaction strength $\theta$. For each value
of $\theta$, the noise strength value is that at which the  SNR is
maximal. The periodicity of $L_\mathrm{coh}(t)$ is clearly
demonstrated. Its amplitude is only slightly dependent on the
interaction strength. Its Fourier component at the fundamental
driving frequency is precisely the numerator of $R_\mathrm{out}$.
Their peak values are not much different from those obtained in a
single bistable unit driven by the same force at the same noise
strength.

In  the right panel of Fig. \ref{fig:3-4}, the behaviors of the
corresponding incoherent parts are depicted. It is remarkable the
fast decay of the fluctuations as well as their small values. As the
denominator of the SNR is the Fourier component of those decaying
functions at the driving frequency, it is clear that those
contributions are small. Consequently, the values of the SNR are
expected to be much enhanced with respect to those values typical of
SR in the linear regime. The enhancement of SR effects in arrays of
interacting bistable units with respect to those in individual units
is then basically a consequence of the strong reduction of the
fluctuation level with respect to the one found in an single driven
unit.

\begin{figure}
\resizebox{0.4\textwidth}{!}
  {\includegraphics{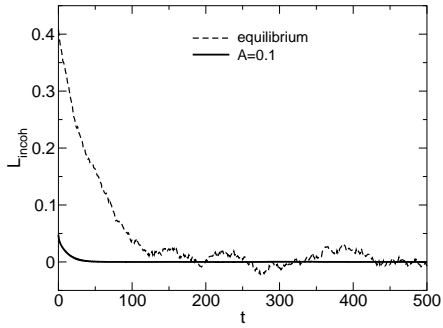}}
\caption{Solid line: The incoherent part of the correlation function
of the global variable for an array of $N=10$ bistable units driven
by a rectangular driving force with amplitude $A=0.1$ and
fundamental frequency $\Omega=0.01$. Dashed line: the equilibrium time correlation function
of the global variable for the same array in the absence of driving. In both cases the coupling parameter is
$\theta=0.5$ and the noise strength value $D=0.16$.}
\label{fig:5}       
\end{figure}

It should also be noted in  the lower panel of  Fig. \ref{fig:1-2}
that the gain can be larger than unity for some ranges of noise
strength values. This feature is a clear indication that the SR
phenomenon observed in the array for the parameter values considered
can not be described within the limits of a linear response theory.
To further understand why linear response theory fails in the cases
considered here, it seems useful to compare the behavior of the
incoherent part of the one-time correlation function of the global
variable and that of the equilibrium correlation function of the
same global variable in an un-driven system. An example of such
comparison is depicted in Fig. \ref{fig:5} for a coupling strength
$\theta=0.5$. The graph clearly indicates that the equilibrium
fluctuations are much larger and longer lasting than the
fluctuations about the average behavior in the driven system. In the
linear response theory description of SR, (see, for instance,
\cite{RMP} and references therein),  it is assumed that the
correlation function of the fluctuations around the average behavior
in a driven system can be safely approximated by their corresponding
equilibrium values in a un-driven one. This is clearly not the case
for the the system at hand.  A detailed study of the validity
conditions of linear response theory can be found in
\cite{C-P5,C-P6}.

\begin{figure}
\resizebox{0.4\textwidth}{!}
  {\includegraphics{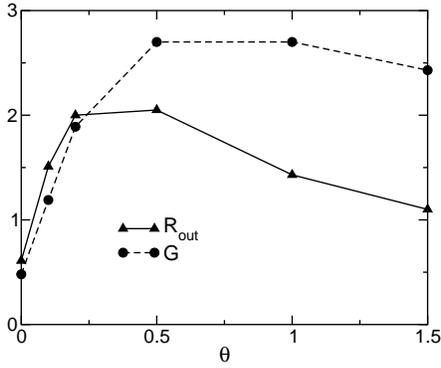}}
\caption{The peak values of $R_\mathrm{out}$  and $G$ vs. the
coupling strength $\theta$ for an array of $N=10$ elements driven by
a rectangular force with $A=0.1$ and fundamental frequency
$\Omega=0.01$. }
\label{fig:6-7}       
\end{figure}

The peak values of $R_\mathrm{out}$ also show a non-monotonic
behavior with $\theta$ as depicted in Fig. \ref{fig:6-7}. As
$\theta$ is raised from zero up to $0.5$, there is an increase in
the peak of the SNR value. This is due to the combination of two
effects: i) the faster decay of the correlation function as $\theta$
is increased (see right panel in Fig.  \ref{fig:3-4}) with the
subsequent decrease of the denominator in the SNR, and ii) the
larger amplitude of the coherent part (see left panel in Fig.
\ref{fig:3-4}). On the other hand, as the $\theta$ values is further
increased, the contribution of the coherent part decreases, while
that of the incoherent part increases and, consequently, the SNR
decreases as the coupling term increases. The gain also shows a
non-monotonic behavior with $\theta$ as depicted in Fig.
\ref{fig:6-7}.

The complexity of the $N$-dimensional potential surface where the
interacting particles move renders the task to give a simple
explanation of the observed non-monotonic behavior with $\theta$ a
difficult one. For an $N$-dimensional surface in the absence of
interactions and driving forces, the whole surface is symmetrical
about the origin with barriers of equal heights along each axis. The
observed non-monotonic behavior can be rationalized in terms of the
periodic rocking of the bistable potential independently along each
axis. Each particle jumps over its corresponding barrier
independently of the other particles, under the influence of the
driving force and the noise. The central limit theorem can be safely
used for independent subunits, so that the increase in the SNR
values is just a size effect.

On the other hand, for coupled systems, the individual stochastic
processes $x_i(t)$ are no longer independent and the central limit
theorem alone is not enough to understand the reported results. The
deformation of the potential surface due to the external driving and
the interaction term might very well reduce the height of the
barriers, eliminate some of them and alter the location of the
minima. Then, one can not rule out the possibility of the existence
of new paths facilitating the transitions between the attractors.
This being the case, a reduction of the fluctuation levels besides
the one coming from the system size is to be expected. For a fixed
small driving amplitude, the amount of distortion of the potential
relief must depend on the strength of the coupling term $\theta$.
For very small values of $\theta$, particles located beyond nearest
neighbor positions are expected to be weakly correlated, with the
correlation length increasing as $\theta$ increases.  One might
expect an increase on the SNR values as $\theta$ increases from
zero, as the incoherent part of the correlation function basically
decreases. This is due to the increasing easiness of transitions
between attractors along the new paths. As $\theta$ is further
increased, the distortion of the potential relief will increase, but
at the same time, most of the particles along the chain will start
to be strongly correlated. In other words, the motion of the
collective variable will resemble more and more the motion of a
particle on a single bistable potential. The chain will behave more
and more like a rigid object. Jumps over the barriers become
increasingly more difficult and this leads to a decrease of the SNR
values. There must be an intermediate interaction strength value so
that a maximum SNR of the output is achieved.

\begin{figure}
\resizebox{0.4\textwidth}{!}
  {\includegraphics{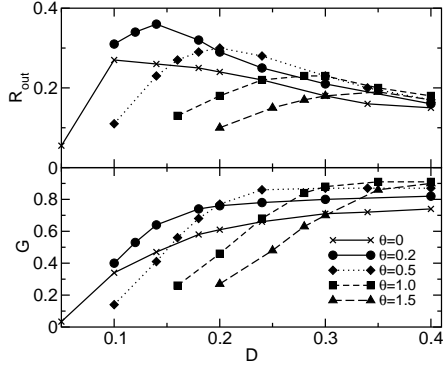}}
\caption{The behavior of $R_\mathrm{out}$  and  $G$ with $D$ for
several
  values of the coupling strength
for an array of $N=10$, driven by a sinusoidal force with $A=0.1$
and frequency $\Omega=0.01$. }
\label{fig:8}       
\end{figure}

It is interesting to study what happens in the case of a sinusoidal
driving force with the same amplitude $A=0.1$ and
 frequency $\Omega=0.01$. In Fig.\ \ref{fig:8} we depict
 $R_\mathrm{out}$ (lower panel) and $G$ (upper panel) vs. $D$ for
 several values of the coupling strength. As it can be seen, the SNR
 values are substantially smaller than the ones observed for a
 rectangular driving force (compare with  Fig.\ \ref{fig:1-2}). The gain
 is always below unity by contrast with the rectangular driving force,
 where the gain can reach values larger than 1.

\begin{figure}
\resizebox{0.4\textwidth}{!}
  {\includegraphics{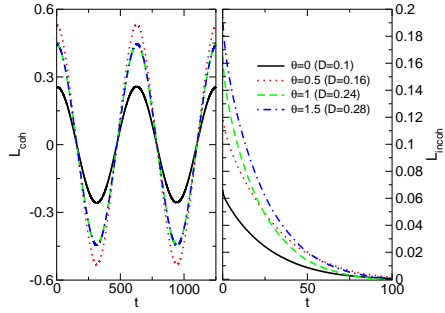}}
\caption{The time behavior of $L_\mathrm{coh}$  and
$L_\mathrm{incoh}$  for several
  values of the coupling strength
for an array of $N=10$, driven by a sinusoidal force with $A=0.1$
and frequency $\Omega=0.01$. }
\label{fig:9}       
\end{figure}

\begin{figure}
\resizebox{0.4\textwidth}{!}
  {\includegraphics{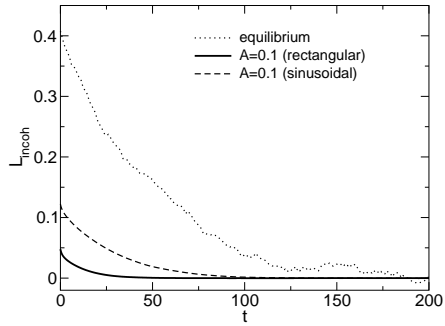}}
\caption{Comparison of the time behavior of $L_\mathrm{incoh}$ in an
array of $N=10$ at equilibrium or driven by either a sinusoidal
force with $A=0.1$ and frequency $\Omega=0.01$, or a rectangular one
with the same amplitude and fundamental frequency. Other parameter
values: $\theta=0.5; D=0.16$}
\label{fig:10}       
\end{figure}

In Fig.\ \ref{fig:9} we show the time dependence of the coherent and
incoherent parts of the correlation function for an array of $N=10$
particles driven by the sinusoidal force. By comparison with  Fig.\
\ref{fig:3-4} we see that the reason why the SNR values are much
smaller for a sinusoidal driving than for a rectangular one is
mainly that $L_\mathrm{incoh}(t)$ is larger in the single frequency
case than in the multi-frequency one. It is not correct to conclude
that because the gain in the case of sinusoidal driving is less than
unity, a linear response theory description is adequate. As shown in
Fig.\ \ref{fig:10}, $L_\mathrm{incoh}(t)$ for a sinusoidal driving
is very different from the equilibrium correlation function in an
un-driven system.

A non-monotonic behavior of the SNR with the coupling constant
$\theta$ for the sinusoidal driving also exists. The qualitative
explanation given before for the rectangular driving still stands.
Note, nonetheless, that the rocking of the multi-dimensional energy
surface brought up by the sinusoidal driving is less drastic than
the one produced by the rectangular one. Even though the forces have
the same amplitude and fundamental frequency, the sinusoidal force
introduces a bias in the energy relief with respect to that in the
zero force case which is continuously changing with time. On the
other hand, the rectangular signal keeps the surface biased most of
the time, except during its instantaneous changes of value. The
system has ample time to relax during those biased intervals and
this explains why the fluctuations are so drastically reduced in the
case of a rectangular driving.

\section{Conclusion}
\label{sec:3} In this work, we have explored the phenomenon of SR in
finite arrays of noisy bistable units with nearest neighbor
coupling, driven by time periodic forces of weak amplitude. Rather
than analyzing the modification of a single unit behavior with
respect to the one in the absence of any interaction, we have
focused our attention in a collective variable describing the
dynamics of the array as a whole.

A strong enhancement in the SR effects associated with the
collective variable with respect to the one found in a system formed
by a single unit has been demonstrated by means of numerical
simulations. In particular, we have shown that, even for rather weak
input amplitudes, the SNR has a non-monotonic behavior with the
noise strength. The different values of the SNR quantifier are much
larger than the corresponding ones found in a single unit system.
Furthermore, the SR gain reaches values higher than unity for some
values of the parameters, indicating that the array is operating in
a nonlinear regime well beyond the limits of the regimes described
by linear response theory. This is to some extent surprising as, for
the driving amplitudes and frequencies considered, the response of a
single unit system is well described by the linear response theory.

There are in principle two main reasons for the enhanced effects
reported here. On the one hand, based on the central limit theorem,
one can expect a decrease in the fluctuation levels with respect to
those found in single unit systems simply because of the system
size. On the other hand, as we have demonstrated in our numerical
simulations, the size effect mechanism is not enough to explain the
reported behavior. The presence of coupling terms between the
subsystems is an essential ingredient to obtain gains larger than
unity.

We believe that the main reason for the strong enhancement observed
is due to the drastic reduction of the value of the incoherent
correlation function and on its correlation time induced by the
system size, the external driving and the coupling term between the
different subunits. Indeed, the comparison of the incoherent
correlation function with the equilibrium correlation function in
finite systems with the same size and coupling parameter, shows the
relevance of the driving force.

An enhancement of the SR effects in finite size arrays with global
coupling (mean field coupling) have also been reported by us. One of
the goals of the present work is to demonstrate the robustness of
our results, regardless of the topology of the connections between
the different subunits. Indeed, a comparison of the numerical
results reported here with those in \cite{us08} indicate that, at
least for small size arrays ($N=10$), the $R_\mathrm{out}$ and gain
values are not much different.

We have also noted that there exists a non-monotonic behavior of the
SNR with respect to the coupling parameter $\theta$, besides the
usual non-monotonic behavior with the noise strength. There are some
indications that non-monotonic behaviors with respect to the system
size might also exist.  Indeed, system size resonances have been
analyzed by other groups for systems of \emph{globally} coupled
nonlinear oscillators using cumulant expansion techniques
\cite{Pikovsky} or nonequilibrium potentials \cite{wio}. Within a
linear response theory description, they report system size
resonance effects. We are presently investigating the issue of the
dependence on the system size in coupled arrays (global and local
coupling) for parameter regimes well beyond the linear response
theory limits.



%
%
%
%

\end{document}